# ENERGY EFFICIENT NEIGHBOUR SELECTION FOR FLAT WIRELESS SENSOR NETWORKS

Saraswati Mishra[1] and Prabhjot Kaur[2]

Department of Electrical, Electronics and Communication Engineering, ITM University, Gurgaon, India
[1]saraswati12ecp020@itmindia.edu
[2]prabhjotkaur@itmindia.edu

## ABSTRACT

*In this paper we have analyzed energy efficient neighbour selection algorithms for routing in wireless sensor networks. Since energy saving or consumption is an important aspect of wireless sensor networks, its precise usage is highly desirable both for the faithful performance of network and to increase the network life time. For this work, we have considered a flat network topology where every node has the same responsibility and capability. We have compared two energy efficient algorithms and analyzed their performances with increase in number of nodes, time rounds and node failures.*

## 1. INTRODUCTION

Wireless sensor networks consist of number of small nodes deployed in an area under supervision. Each node has limited storage, computational and sensing capability and limited energy resource, as nodes are battery operated. Since energy is the main concern in wireless sensor networks (WSN) to maximize the performance and to increase the lifetime of a network, various approaches are implemented to reduce energy consumption in a network. Most of the energy is consumed during idle period and during transmission of data from one node to another i.e. routing. An efficient media access control (MAC) and routing protocol should be designed to save energy. While MAC protocol targets at reduction of energy in scanning and accessing the channel, routing protocol helps to reduce the energy requirement for end-to-end transmission.

In WSN, there are number of routing protocols that have been proposed for different network criteria. Based on the network topology WSN protocols have been categorized as – flat network protocols and hierarchical protocols as shown in Figure 1. Protocols that fall under hierarchical class select one head amongst all and form a hierarchy [1]. This hierarchy may be a cluster, a chain or a grid. Cluster head or a leader collects the information from all the other nodes in its region and sends it to the sink node or gateway node. Some examples are low energy adaptive clustering hierarchy (LEACH), power efficient gathering in sensor information systems (PEGASIS), virtual grid architecture (VGA), etc. The work presented in this paper considers flat protocols and thus we are not including descriptions of hierarchical protocols and will only be focusing on flat protocol strategies in the rest of this paper. In a flat network, every node is treated equally in terms of responsibility and capability. There is no master and no slave. Flat network protocols are further classified into – quality of service (QoS) based protocols, data centric protocols and location based protocols. Some examples of this type of protocols are sensor protocol for information via negotiation (SPIN), directed diffusion (DD), gradient based routing (GBR) and geographic and energy aware routing (GEAR) [2].

Most of the protocols mentioned above implement energy saving as an important feature and accordingly, before delivering the data packet to the next hop neighbour, the source or intermediate node checks residual energy or consumed energy to decide the neighbour for the data forwarding. Protocols working on this approach are known as energy centric or energy aware protocols. In data centric protocols, sink sends queries and waits for data. Attribute-based naming is necessary to specify the property of data that data can be requested through queries. QoS based routing is different from address based routing mechanism used in data centric protocols. It selects the path based on some previous knowledge of resource availability and maximum tolerable delay as well as QoS requirement of a network. Also for optimum routing it adaptively allocates the available resources to maintain QoS. Location based protocols are used for routing queries towards targeted region of sensor network. Location information of next hop neighbour should be known to each sensor node. This information is used to calculate the distance between two nodes so that energy consumption can be determined [7]. Out of these techniques, we have considered data centric approach for the analysis and analyzed two different strategies to reduce energy consumption during routing. The detail description of these techniques is given in the next section.

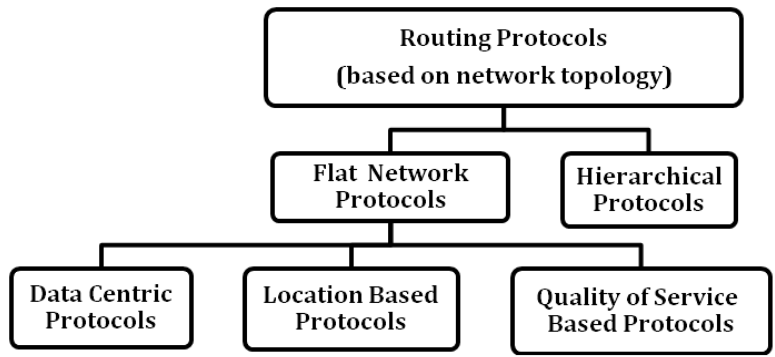

Figure 1. Classification of routing protocols based on network topology

The rest of this paper is organized as follows. Section 2 briefs the routing algorithms considered for analysis. Section 3 presents different approaches that have been implemented on the chosen routing protocol to attain energy efficiency. The simulation setup and results are discussed in section 4 and the paper concludes with section 5.

## 2. DATA CENTRIC PROTOCOLS

As we have discussed, the data centric protocols work in a flat networks. The working principle of these routing protocols is based on query (or a request) [4]. Query may be generated either by a sink node or source node.

In first case, sink broadcasts query to get specific type of data, any node having that specific data replies back. In second case, source sends the signal to specify that it is having some specific data, interested node can receive that by sending request. As we can observe that the routing is taking place via negotiation, it is important here to mention that the negotiation based protocols are the special class of data centric protocols. Negotiation based protocols may be of two stages – query and data or it may be of three stages – metadata, query and data. Metadata is a packet that contains information regarding the data of the node. Format of metadata may vary with the variation in application. Traditional flat protocols like flooding and gossiping have various drawbacks and limitations like implosion, data redundancy and resource blindness [3]. These can be overcome by use of data centric protocols. According to the stages best example of three stages negotiation based protocol is SPIN and example of two stage protocol is DD. Their brief description is given below.

## 2.1. Sensor Protocol for Information via Negotiation

SPIN is a data dissemination protocol that disseminates its information to all the nodes in its vicinity. This protocol works in three stages. First the node having data sends the advertising message (ADV) to the single hop neighbour. ADV acts as a metadata here. The interesting neighbour replies with request message REQ to indicate that it needs the data and finally the data is sent to requesting node. SPIN is classified in different classes like point-to-point (SPIN-PP), broadcast (SPIN-BC) reliable (SPIN-RL) and energy centric (SPIN-EC) and are used depending upon the application [4].

## 2.2. Directed Diffusion

The DD is again a flat network protocol that works on a principle of flooding. Here for a need of specific data sink node floods the interest signal in the network through the neighbours. After receiving a request every node maintains an interest cache. This is maintained till the gradient is not formed. The gradient is a reply link through which a request was received. Gradient contains all the information about the path i.e. data rate, duration etc. among all the paths formed from sink to source the best path is selected through the reinforcement process that means data is sent through selected shortest path and hence prevents further flooding [4].

From above discussed protocols we have analyzed three stage negotiation based protocol with the addition of subroutine that makes it energy efficient.

## 3. ENERGY EFFICIENT APPROACH

There are various approaches to minimize the energy consumed by the routing protocol. To make flat routing protocols more energy efficient - 1) we can select the neighbour which is closest to base station so that the number of hops to perform routing is minimum, this will save energy 2) another approach is to select the neighbour or next hop having maximum residual energy among all and 3) select the path towards the destination or sink that consumes minimum energy. Among all the approaches mentioned above we have applied second and third approach for the analysis. Detailed specification is given below.

## 3.1. Selection of Neighbour having Highest Energy

In this selection approach, when source want to transmit data to the destination which is multiple hop away from the source then the source checks the energy level of all the neighbours and selects the one having highest energy among all. Similarly, all intermediate nodes find out the neighbour with highest energy and deliver the data to that node and finally the data packet reaches to the destination. In the rest of this paper, we have referred this technique as highest energy (HE).

## 3.2. Selection of Path that Consumes Minimum Energy

In this process the source node or sending node at first, estimates the total energy that will be consumed by all possible paths formed in multipath communication scenario and then selects the best path toward base station which will consume minimum energy amongst all paths during transmission of data packets. This technique is referred as minimum energy consumption route (MECRT).

## 4. SIMULATION SET UP AND ANALYSIS

A flat sensor network was created and above said routing protocols were compared using SENSORIA – a Graphical User Interface (GUI) based simulator [6]. The details of simulation setup are given below in table1.

Routing protocol selected is a negotiation based protocol that works on three stages as we have discussed earlier (replica of SPIN). Simulation takes place till there is a path (nodes are alive) between source and destination to forward packets, otherwise it gets terminated. The nodes are randomly deployed and are dynamic in nature.

Table 1. Simulation parameters and values

| Parameters | Values |
| --- | --- |
| Number of nodes | 50 |
| Energy / node | 0.5 J (homogenous) |
| Simulation area | 50m X 50m |
| Transmission range of each node | 15 m |
| Sensing range (each node) | 8m |
| Location of base station | 25m X 150 m |
| Data packet | 2000 bits |
| Control packet | 248 bits |
| Data transmission speed | 100 bits/sec. |
| Bandwidth | 5000 bits/sec. |

### 4.1. Analysis and comparison of routing protocols for their energy consumption

First, we have studied the impact of energy awareness on routing protocol. Selected protocol has been made energy efficient by the application of minimum energy consuming path selection algorithm, MECRT. The comparison of this protocol with its counterpart i.e. the technique that does not account energy consumption of the path during the transmission is done. The result shows that the energy saving capability of a network is more in which MECRT is implemented, as shown in figure 2. Life time of a network is also increased by the application of MECRT as compared to the life of a network that works without MECRT routing protocol. We can also see from the graph that energy consumption is less in MECRT hence it works for comparatively long time rounds.

Further, we have compared the performance of HE and MECRT that makes routing more energy efficient. Highest Energy neighbour selection and MECRT are applied on three stage negotiation based protocol like SPIN, where communication takes place through the exchange of metadata, query (or request) and then data, as we discussed earlier. These energy aware techniques used in our scenario helped in increasing life time and its performance.

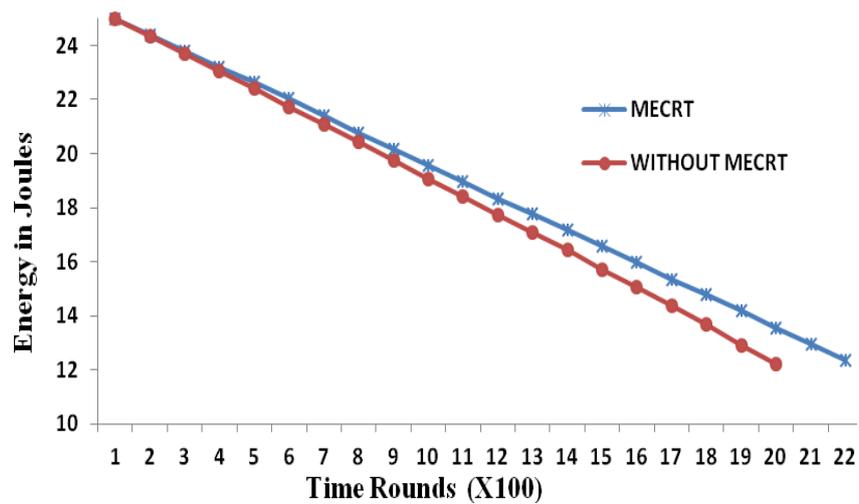

Figure 2. Comparison of algorithms (to select the route among all the available routes) with and without energy consideration

Both the protocols were applied on the same network with same parameters as described above and their performances were recorded. The performance analysis represents that MECRT gives better results than HE in terms of reduction of energy consumed by the network during routing as shown in Figure 3. Moreover when network route was discovered using HE algorithm, its life time was shortened as compared to the network that has MECRT as energy saving mechanism. Nodes death rate frequently increased in case of HE after certain time round compared to the MECRT algorithm which is shown in Figure 4. Both plots represent the energy decay and node failure with respect to the time round.

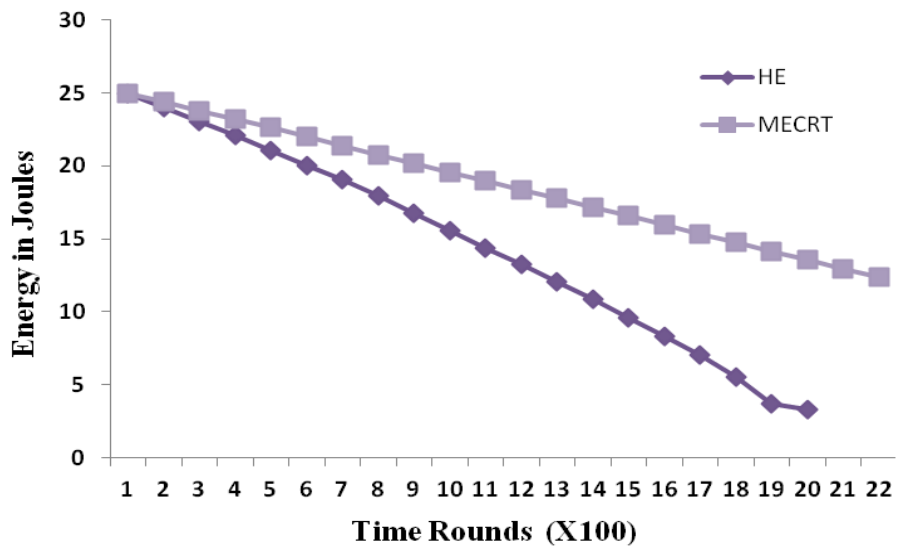

Figure 3. Comparison of algorithms HE and MECRT in terms of energy degradation of a network

In small network size, i.e. with few numbers of nodes, the difference was not significant. As flat routing protocols are mainly designed for small and medium size networks, we have simulated their performance on networks having nodes ranging from 10 to 200.

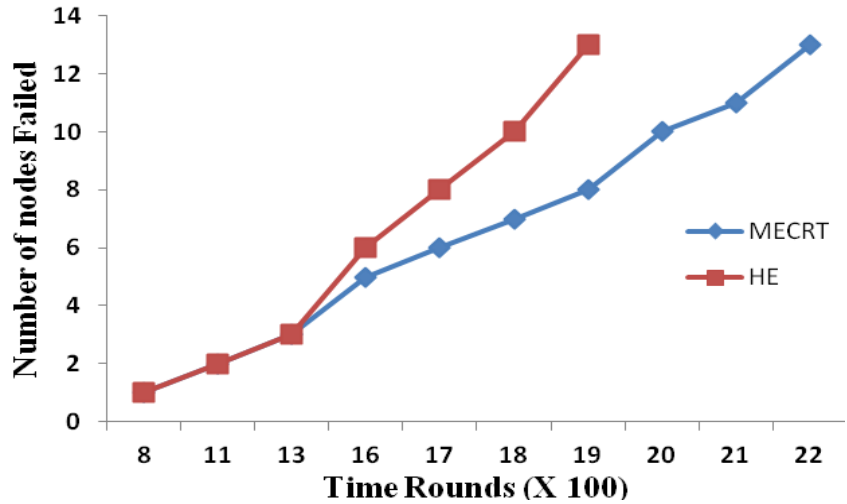

Figure 4. Comparison of algorithms HE and MECRT in terms of number of nodes failure

As the number of nodes increased to 25 to 150 we can easily identify the difference and can conclude that MECRT gives better results than HE for the static network topology in terms of energy consumption as well as network life time. The comparison with increase in number of nodes in a scenario is shown in figure 5.

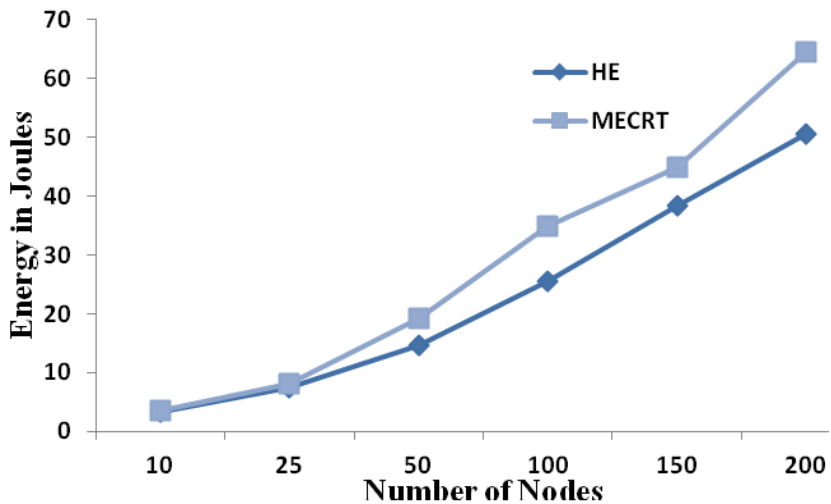

Figure 5. Comparison of algorithms HE and MECRT when number of nodes are increased

## 5. CONCLUSIONS

Through this paper, it is clear that energy efficient routing protocols helps to save energy in wireless sensor network and should be used in scenarios where energy consumption of sensors is a constraint. We have compared HE and MECRT in a flat network topology, to reduce energy consumption during routing. In a network that contains 10 or 20 nodes, any approach either HE or MECRT will give almost similar result. Hence any of the algorithms can be implemented in a routing protocol. Through the experimental analysis we can conclude that MECRT is better for medium to large network size, where node selects a path that consumes minimum energy among all available paths for data forwarding as compared to the HE algorithm where node delivers the data to the neighbouring node having highest energy. However, both these techniques do not guarantee the shortest route selection or fast routing mechanism. These protocols only deal with less expenditure of energy during routing. Hence more effective routing algorithm can be designed in future that will tend to select shortest path while assuring least energy consumption.

## ACKNOWLEDGEMENTS

We would like to thank G. Al.-Mashaqbeh for providing permission and guidance to install and understand SENSORIA simulator.

**Authors**

Saraswati Mishra is pursuing post graduation in Electronics and Communications from Institute of Technology and Management University, Gurgaon, India. She did her graduation in Electronics and Software Technology from LAD College, Nagpur University, Nagpur, Maharashtra, India. She is currently working on Cognitive radio technology.

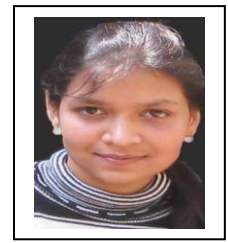

Prabhjot Kaur did her Ph.D. on MAC models for Dynamic Spectrum Access in Cognitive Radio Networks from National Institute of Technology, Jalandhar, and her Masters of Engineering from Punjab University, Chandigarh India. She is currently working as Associate Professor and Deputy Dean (RDIL) with ITM University, Gurgaon, India. Her research interests include dynamic spectrum allocation, Ad-hoc Networks, Green Networks, MIMO, software defined radios and Cognitive Radios.

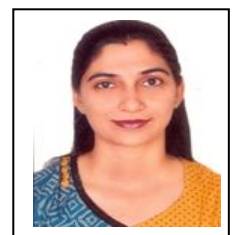

She has completed a research project funded under Research Grant from AICTE, Govt. of India under research promotion scheme in April 2010 and an international travel grant for attending IEEE conference by Department of Science and Technology, Govt. of India. She received the `Best Emerging Researcher Award` of the year 2012 at ITM University, India and Best Paper recognition at an International conference, Malaysia. She is member, IEEE and life member of IETE and ISTE societies.